\documentclass[twocolumn,5p,number,letterpaper]{elsarticle}%
\usepackage{amsmath}
\usepackage{epsfig}
\usepackage{natbib}
\usepackage{accents}
\usepackage{amsfonts}
\usepackage{amssymb}
\usepackage{color}
\usepackage{subfig}
\usepackage{graphicx}%
\providecommand{\U}[1]{\protect\rule{.1in}{.1in}}
\providecommand{\U}[1]{\protect\rule{.1in}{.1in}}

\journal{Solid State Communications}
\begin{document}
\begin{frontmatter}
\title{Actuation, propagation, and detection of transverse magnetoelastic waves in ferromagnets}
\author{Akashdeep Kamra\fnref{a}}
\author{Gerrit E. W. Bauer\fnref{a,b}}
\address[a]{Kavli Institute of NanoScience, Delft University of Technology, Lorentzweg 1, 2628 CJ Delft, The Netherlands}
\address[b]{Institute for Materials Research and WPI-AIMR, Tohoku University, Sendai 980-8577, Japan}
\begin{abstract}
We study propagation of ultrasonic waves through a ferromagnetic medium with special attention to the boundary conditions at the interface with an ultrasonic actuator. In analogy to charge and spin transport in conductors, we formulate the energy transport through the system as a scattering problem. We find that the magneto-elastic coupling leads to a non-vanishing magnetic (elastic) energy accompanying the acoustic (spin) waves with a resonantly enhanced effect around the anti-crossing in the dispersion relations. We demonstrate the physics of excitatinig of magnetization dynamics via acoustic waves injected around the ferromagnetic resonance frequency.
\end{abstract}
\end{frontmatter}

\section{Introduction}

While the exchange interaction is the largest energy scale of ferromagnets,
explaining Curie temperatures of up to 1000 K, the static equilibrium and
dynamic properties of the magnetization field in ferromagnetic materials are
governed by the dipolar and crystal anisotropy fields \cite{textbook}. Since
the total angular momentum of an isolated system is conserved, any change of
the magnetization exerts a torque on the underlying lattice, as measured
by Einstein and de Haas \cite{Einstein}. Vice versa, a rotating lattice can magnetize a demagnetized ferromagnet \cite{Barnett}. The coupled equations of motion of
lattice and magnetization fields have been treated in a seminal paper reported by
Kittel \cite{kittelpr}. The magnetoelastic coupling parameters are material
constants well known for many ferromagnets\ \cite{chikazumi}.

Interest in magnetoelastic coupling has recently been revived in the context
of the \textquotedblleft spin mechanics\textquotedblright\ concept covered by
the present special issue [Editorial SSC]. Here we are interested in the
magnetization dynamics acoustically induced by injecting ultrasound into
ferromagnets by piezoelectric actuators as bulk \cite{Uchida} or surface
\cite{Goennenwein} plane acoustic waves. The magnetization dynamics in these
experiments is conveniently detected by spin pumping \cite{pump} into a normal
metal that generates a voltage signal via the inverse spin Hall effect
\cite{Saitoh}.

Some of the consequences of magnetoelastic coupling have already been
investigated theoretically \cite{kittelpr,Kobayashi,Dreher} and experimentally
\cite{Boemmel,Feng} in the literature. While the coupled magnetoelastic dynamics
has been well understood decades ago \cite{kittelpr,Kobayashi,Tiersten}, much
less attention has been devoted to the interfaces that are essential in order
to understand modern experiments on nanostructures and ultrathin films. The
Landauer$-$B\"{u}ttiker electron transport formalism based on scattering theory
is well suited to handle these issues thereby helping to understand many
problems in mesoscopic quantum transport and spintronics
\cite{Datta,Nazarovblanter}. Here we formulate scattering theory of lattice
and magnetization waves in ferromagnets with significant magnetoelastic
coupling. Rather than attempting to describe concrete experiments, we wish to
illustrate here the usefulness of this formalism for angular momentum and
energy transport.

We consider magnetization dynamics actuated by ultrasound for the simplest
possible configuration in which the magnetization direction is parallel to
the wave vector of sound with transverse polarization (shear waves). The
corresponding bulk propagation of magnetoelastic waves was treated long ago by
Kittel \cite{kittelpr} who demonstrated that the axial symmetry reduces the
problem to a quadratic equation. The injected acoustic energy is partially
transformed into magnetic energy by the magnetoelastic coupling that can be
detected by spin pumping into a thin Pt layer. For this symmetric
configuration and to leading order a purely AC voltage is induced by the
inverse spin Hall effect (ISHE) \cite{Hujun} that might be easier to observe
by acoustically induced rather than rf radiation induced spin pumping, since
in the former the Pt layer is not directly subjected to electromagnetic
radiation. The configuration considered by Uchida \textit{et al}.
\cite{Uchida}, in which pressure waves generate a DC ISHE\ voltage by a
magnetization parallel to the interfaces, will be discussed elsewhere.

\section{Kittel's equations}

We consider a ferromagnet with magnetization texture $\mathbf{M}\left(
\mathbf{r},t\right)  $ with constant saturation magnetization $\left\vert
\mathbf{M}\right\vert =M_{0}$. In the following we consider small fluctuations
around the equilibrium magnetization $M_{0}\mathbf{z.}$ The classical
Hamiltonian can be written as the sum of different energies%
\begin{equation}
\mathcal{H=H}_{Z}+\mathcal{H}_{ex}+\mathcal{H}_{me}+\mathcal{H}_{p}
\label{Htot}%
\end{equation}
The magnetic Zeeman energy reads
\begin{equation}
\mathcal{H}_{Z}=\frac{\omega_{0}}{2\gamma M_{0}}\left(  M_{x}^{2}+M_{y}%
^{2}\right)  \label{ex}%
\end{equation}
where $\gamma=\left\vert \gamma\right\vert $ is the gyromagnetic ratio,
$\omega_{0}=\mu_{0}\gamma H$ is the magnetic resonance frequency for an
effective magnetic field $H\mathbf{z}$ and $\mu_{0}$ the permeability of free
space\textbf{.} The exchange energy cost of the fluctuations
\begin{equation}
\mathcal{H}_{ex}=\frac{A}{M_{0}^{2}}\left[  \left(  \nabla M_{x}\right)
^{2}+\left(  \nabla M_{y}\right)  ^{2}\right]
\end{equation}
where $A$ is the exchange constant. The magnetoelastic energy for cubic
crystals and magnetization in the $z$-direction can be parameterized by the
magnetoelastic coupling constant $b_{2}:$
\begin{equation}
\mathcal{H}_{me}=\frac{b_{2}}{M_{0}}\left(  M_{x}\frac{\partial R_{x}%
}{\partial z}+M_{y}\frac{\partial R_{y}}{\partial z}\right)
\end{equation}
where $\mathbf{R}=\left(  R_{x},R_{y},0\right)  $ is the displacement vector
of a transverse lattice wave propagating in the $z$-direction. $\mathcal{H}%
_{me}$ can be interpreted as a Zeeman energy associated with a dynamic
transverse magnetic field $b_{2}\partial_{z}\mathbf{R}$. The corresponding
elastic energy reads
\begin{equation}
\mathcal{H}_{p}=\frac{\rho}{2}\mathbf{\dot{R}}^{2}+\frac{\alpha}{2}\left[
\left(  \frac{\partial R_{x}}{\partial z}\right)  ^{2}+\left(  \frac{\partial
R_{y}}{\partial z}\right)  ^{2}\right]
\end{equation}
in terms of the mass density $\rho$ and shear elastic constant $\alpha.$

The total Hamiltonian $\mathcal{H}$ defines the equations of motion of the
coupled $\mathbf{R}$ and $\mathbf{M}$ fields\textbf{.} The results in momentum
and frequency space $X\left(  t\right)  =x\left(  k,\omega\right)  e^{i\left(
kx-\omega t\right)  }$ can be simplified by introducing circularly polarized
phonon and magnon waves $m^{\pm}=m_{x}+i\sigma m_{y}$, $r^{\pm}=r_{x}+i\sigma
r_{y}$ $\left(  \sigma=\pm1\right)  $, leading to \cite{kittelpr}
\begin{equation}
\left(
\begin{array}
[c]{cc}%
i(\omega-\sigma\omega_{m}) & \sigma\gamma b_{2}k\\
i\frac{b_{2}k}{M_{0}} & \omega^{2}\rho-k^{2}\alpha
\end{array}
\right)  \left(
\begin{array}
[c]{c}%
m^{\sigma}\\
r^{\sigma}%
\end{array}
\right)  =0
\end{equation}
where $\omega_{m}=\omega_{0}+Dk^{2}$ and $D=2A\gamma/M_{0}$ is the spin wave
stiffness. This secular equation is quadratic in $k^{2}$ with 4 roots $\left(
s=\pm1\right)  :$%
\begin{equation}
(k_{s}^{\sigma})^{2}=\frac{\rho\omega^{2}}{2\alpha}-\frac{\omega_{0}%
-\sigma\omega}{2D}+\frac{\gamma b_{2}^{2}}{2\alpha M_{0}}+s\sqrt
{\Delta^{\sigma}} \label{ksq}%
\end{equation}
with discriminants
\begin{equation}
\Delta^{\sigma}=\left(  \frac{\rho\omega^{2}}{2\alpha}-\frac{\omega_{0}%
-\sigma\omega}{2D}+\frac{\gamma b_{2}^{2}}{2\alpha DM_{0}}\right)  ^{2}%
+\frac{\omega^{2}\rho}{\alpha D}\left(  \omega_{0}-\sigma\omega\right)  .
\end{equation}
The corresponding eigenstates are given by the spinor
\begin{equation}
\psi_{s}^{\sigma}\left(  \omega\right)  =\left(
\begin{array}
[c]{c}%
m_{s}^{\sigma}\\
r_{s}^{\sigma}%
\end{array}
\right)  =N_{s}^{\sigma}\left(
\begin{array}
[c]{c}%
M_{0}\\
ib_{2}k_{s}^{\sigma}/\left(  \omega^{2}\rho-\left(  k_{s}^{\sigma}\right)
^{2}\alpha\right)
\end{array}
\right)  ,
\end{equation}
where $N_{s}^{\pm}$ is a dimensionless normalization factor.

\begin{figure}[tbh]
\centering
\subfloat[]{\includegraphics[scale=0.23]{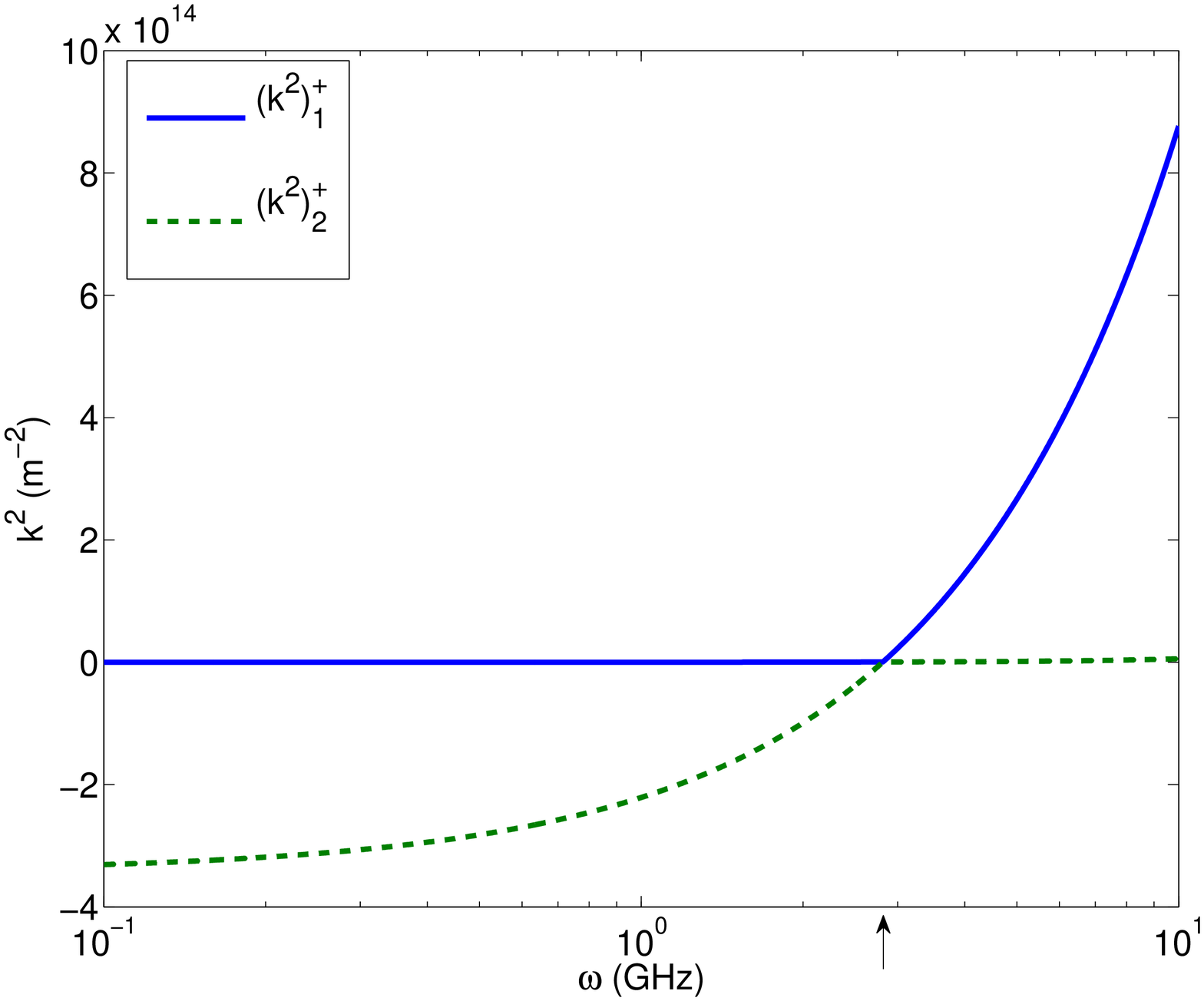}} \newline%
\subfloat[]{\includegraphics[scale=0.23]{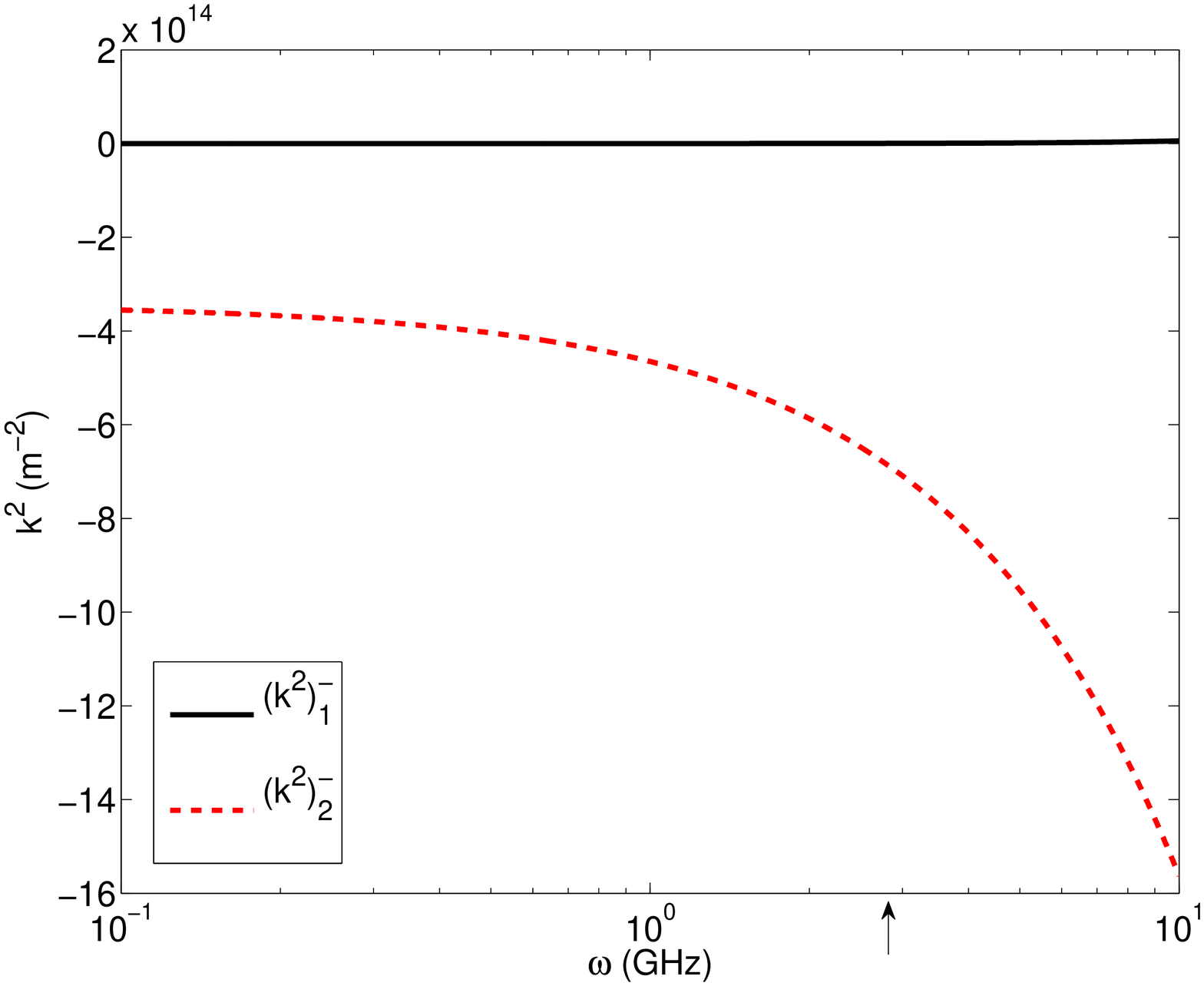}} \newline%
\subfloat[]{\includegraphics[scale=0.23]{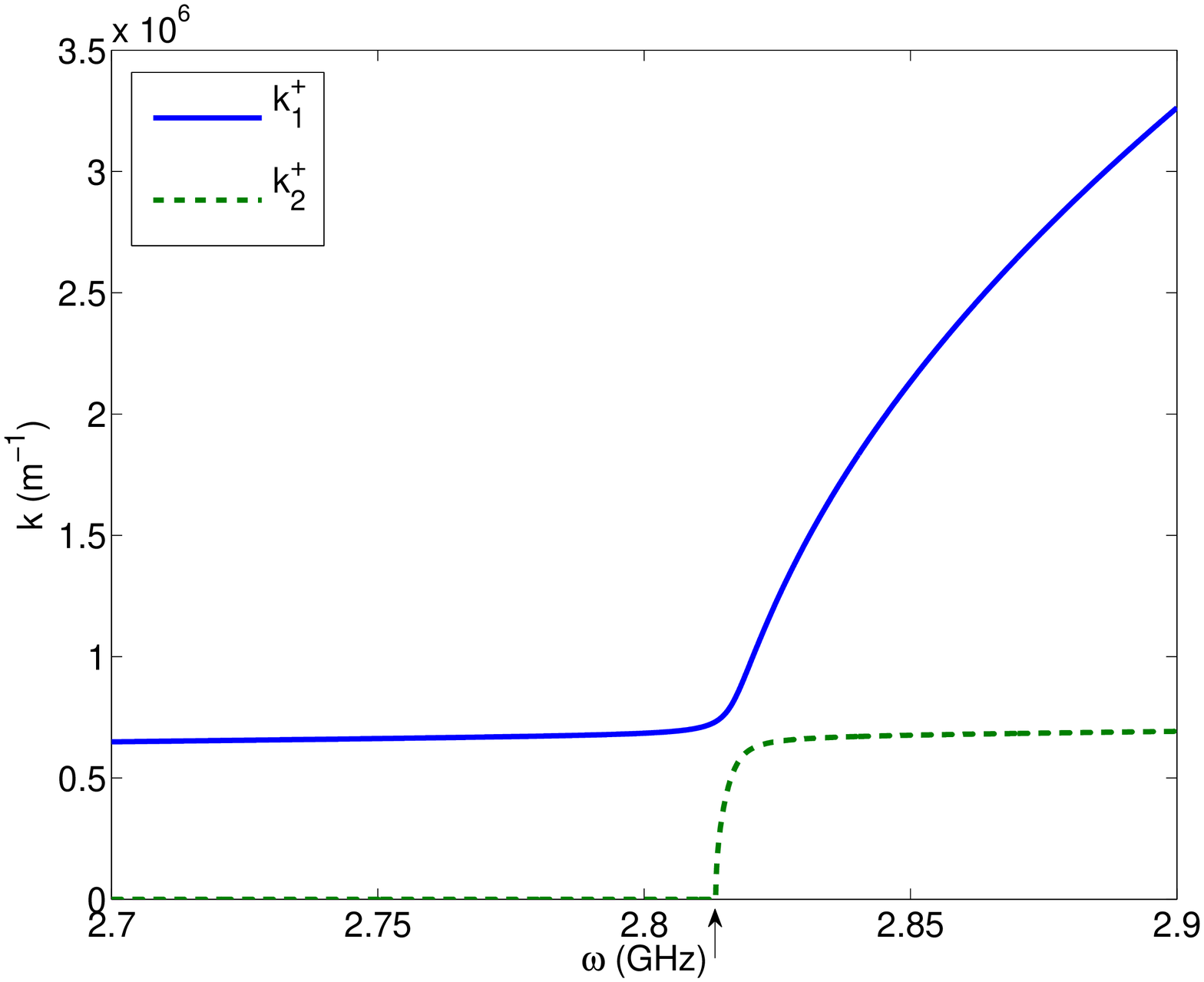}} \newline%
\subfloat[]{\includegraphics[scale=0.23]{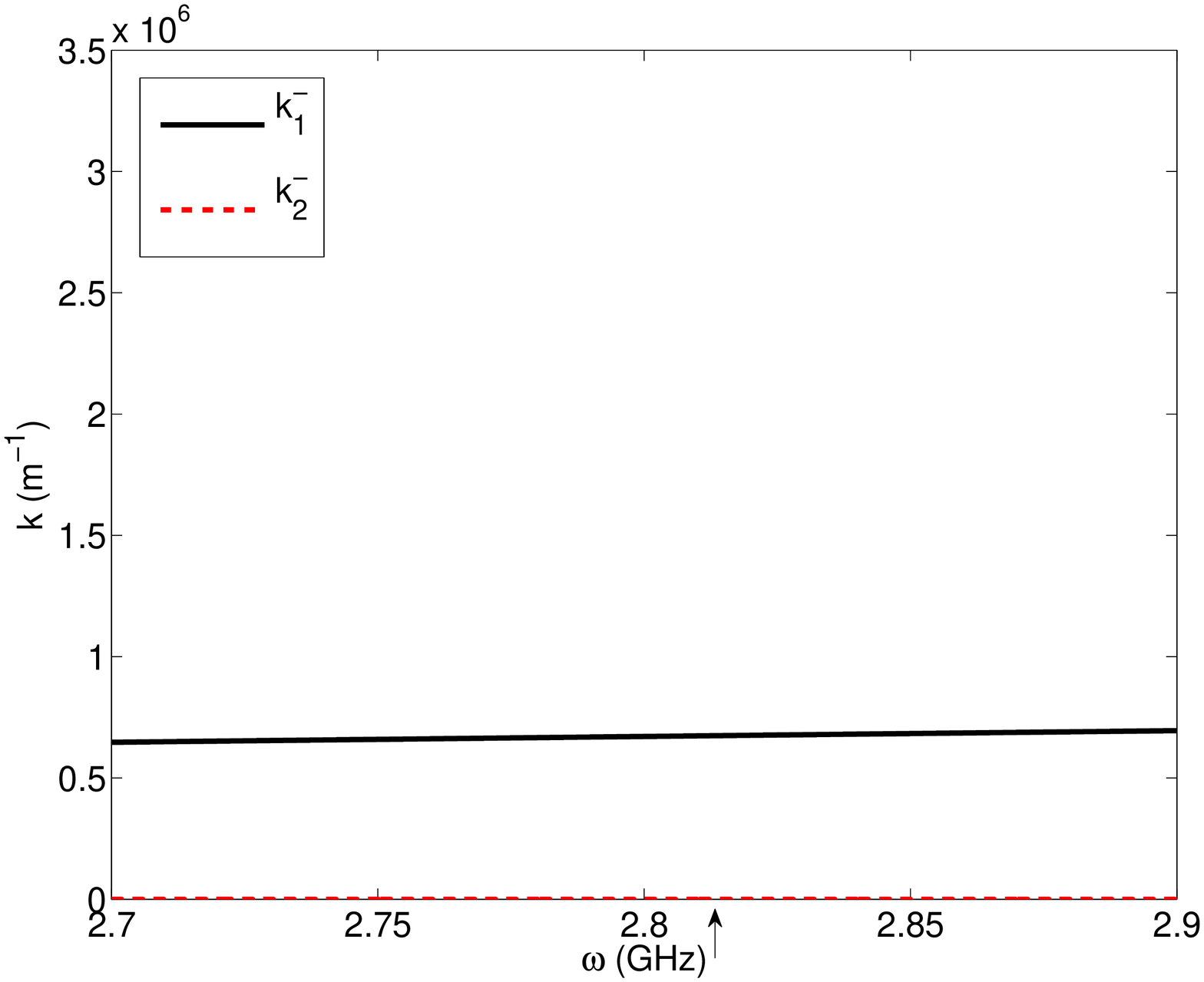}} \caption{Dispersion
relations of magnetoelastic waves in Yttrium Iron Garnets according to Eq.
(\ref{ksq}). (c-d) [(a-b)] depict the [squared] momenta of the eigenstates for
polarization along and against the magnetic order parameter as a function of
frequency. Imaginary momenta appear on the abscissa corresponding to zero real
part. The FMR resonance frequency $\omega_{0}$ is indicated by the arrow on
the abscissa.}%
\label{dispall}%
\end{figure}

The dispersion is plotted in Fig. \ref{dispall} for the parameters appropriate
for Yttrium Iron Garnet (YIG): $M_{0}=1.4\times10^{5}\,\mathrm{A/m}%
,$\thinspace$b_{2}=5.5\times10^{5}\,\mathrm{J/m^{3}},$\thinspace
$H=8\times10^{4}\,\mathrm{A/m},$\thinspace$D=8.2\times10^{-6}\,\mathrm{m^{2}%
/s},$\thinspace$\gamma=2.8\times10^{10}\,\mathrm{Hz\,T}^{-1},$\thinspace
$\rho=5170\,\mathrm{kg/m^{3}},\,\alpha=7.4\times10^{10}\,\mathrm{Pa}$
\cite{para1,para2,para3} and $\mu_{0}=4\pi\times10^{-7}\mathrm{NA}^{-2}$. In
Fig. \ref{dispall}(a) we plot the solutions for waves rotating with the
magnetization that appear to be completely phonon (small dispersion) or magnon
like (large dispersion). The latter are evanescent ($k^{2}<0$) below the spin
wave gap $\omega_{0}.$ The low-frequency anticrossing is better seen in Fig.
\ref{dispall}(c) in which the momentum is plotted on an expanded scale. Spin
waves precessing against the magnetization, $\sigma=-1,$ are always evanescent
and there is no (anti)crossing with the propagating phonons. When the
magnetoelastic coupling is switched off $\left(  b_{2}\rightarrow0\right)  $
and $\alpha>4\rho D\omega_{0}$ the pure lattice wave $\omega_{p}=\sqrt
{\alpha/\rho}k$ and spin wave $\omega_{m}=\omega_{0}+Dk^{2}$ dispersions may
cross twice
\begin{gather}
\omega_{c}=\frac{\alpha}{2\rho D}\left(  1\pm\sqrt{1-\frac{4\rho D\omega_{0}%
}{\alpha}}\right) \\
\overset{4\rho D\omega_{0}\ll\alpha}{=}\left\{
\begin{array}
[c]{c}%
\omega_{0}\\
\frac{\alpha}{\rho D}%
\end{array}
=%
\begin{array}
[c]{c}%
2.8\,\mathrm{GHz}\\
1.7\,\mathrm{THz}%
\end{array}
\right.
\end{gather}
where in the second step we take the limit of small $D$. Around these
degeneracy points, of which only the low frequency one is relevant here, the
effects of the magnon$-$phonon coupling are most pronounced. In the zero
frequency limit the solution with $(k_{s}^{\sigma})^{2}\rightarrow0$
represents a phonon mode with zero wave number. $(k_{s}^{\sigma})^{\pm}=\gamma
b_{2}^{2}/\left(  \alpha M_{0}\right)  -\omega_{0}/D$ is a purely evanescent
magnon for small $b_{2}\ $that in principle may become a real excitation when
the coupling of the lattice is strong enough to overcome the spin wave gap.

\section{Energy flux}

Energy conservation implies $\mathbf{\nabla}\cdot\vec{F}=-\partial
\mathcal{H}/\partial t,$ where the energy flux $\vec{F}=F\mathbf{\hat{z}}$
consists of phonon and magnon contributions. In time and position space
\cite{akhiezer}%
\begin{align}
F\left(  z,t\right)   &  =-\int dz\frac{\partial\mathcal{H}}{\partial
t}=-\frac{2A}{M_{0}^{2}}\frac{\partial M_{x}}{\partial z}\frac{\partial M_{x}%
}{\partial t}\nonumber\\
&  -\left(  \alpha\frac{\partial R_{x}}{\partial z}+\frac{b_{2}}{M_{0}}%
M_{x}\right)  \frac{\partial R_{x}}{\partial t}+\left(  x\longleftrightarrow
y\right)  .
\end{align}
For a plane wave oscillating with frequency $\omega$
\begin{align}
X\left(  z,t\right)   &  =x\left(  z,\omega\right)  e^{-i\omega t}+x^{\ast
}\left(  z,\omega\right)  e^{i\omega t}\\
&  =x\left(  \omega\right)  e^{i\left(  kx-\omega t\right)  }+x^{\ast}\left(
\omega\right)  e^{-i\left(  kx-\omega t\right)  }%
\end{align}
the time-averaged energy flux reads%
\begin{equation}
\bar{F}\left(  z\right)  _{x}=-2\omega\operatorname{Im}\left[  \frac{D}{\gamma
M_{0}}m_{x}\partial_{z}m_{x}^{\ast}+\alpha r_{x}\partial_{z}r_{x}^{\ast}%
+\frac{b_{2}}{M_{0}}r_{x}m_{x}^{\ast}\right]  .
\end{equation}

In the absence of magnetoelastic coupling, pure phonon and magnon waves
\begin{align}
\psi_{s}^{\left(  m\right)  }  &  =N_{s}^{\left(  m\right)  }M_{0}\left(
\begin{array}
[c]{c}%
1\\
0
\end{array}
\right)  e^{i\left(  k_{s}z-\omega t\right)  }\\
\psi_{0}^{\left(  p\right)  }  &  =N_{p}\left(
\begin{array}
[c]{c}%
0\\
k^{-1}%
\end{array}
\right)  e^{i\left(  kz-\omega t\right)  }%
\end{align}
carry, respectively, the energy fluxes $\bar{F}^{\left(  p\right)  }$ and
$\bar{F}_{s}^{\left(  m\right)  }:$
\begin{gather}
\bar{F}^{\left(  p\right)  }=N_{p}^{2}2\alpha\omega/k=N_{p}^{2}2\alpha
\sqrt{\alpha/\rho}\,\\
\bar{F}_{s}^{\left(  m\right)  }=\left(  N_{s}^{\left(  m\right)  }\right)
^{2}\frac{2DM_{0}}{\gamma}\omega k_{s}\\
=\left(  N_{s}^{\left(  m\right)  }\right)  ^{2}\left\{
\begin{array}
[c]{c}%
\frac{2D}{\gamma M_{0}}\omega\sqrt{\left\vert s\omega-\omega_{0}\right\vert
/D}\\
0
\end{array}
\,\text{for }%
\begin{array}
[c]{c}%
s\omega>\omega_{0}\\
s\omega<\omega_{0}%
\end{array}
\right.  .
\end{gather}
One of the magnon states is always evanescent, while the other becomes
propagating for frequencies above the magnon gap $\omega_{0}.$ It is then
convenient to define%
\begin{equation}
\left(  N_{p}\right)  ^{2}=\frac{1}{2\alpha\sqrt{\alpha\rho}};\;\left(
N_{s}^{\left(  m\right)  }\right)  ^{2}=\frac{\gamma}{2\omega M_{0}}%
\frac{\Theta\left(  s\omega-\omega_{0}\right)  }{\sqrt{D\left\vert
s\omega-\omega_{0}\right\vert }}.
\end{equation}
such that each state carries a unit of flux. The flux carried by propagating
$\left(  \operatorname{Im}k_{s}^{\sigma}=0\right)  $ mixed states reads%
\begin{align}
\bar{F}_{s}^{\sigma}  &  =2\omega k_{s}^{\sigma}\left(  N_{s}^{\sigma}\right)
^{2}\left[  \alpha\left(  \frac{b_{2}k_{s}^{\pm}}{\omega^{2}\rho-\left(
k_{s}^{\pm}\right)  ^{2}\alpha}\right)  ^{2}\right. \nonumber\\
&  \left.  -\frac{b_{2}^{2}}{\omega^{2}\rho-\left(  k_{s}^{\pm}\right)
^{2}\alpha}+\frac{DM_{0}}{\gamma}\right]  . \label{prop}%
\end{align}
while the time-averaged flux for evanescent waves with $\operatorname{Im}%
k_{s}^{\sigma}\neq0$ and $\operatorname{Re}k_{s}^{\sigma}=0$ can be shown to
vanish identically. By setting $\bar{F}_{s}^{\sigma}$ to unit flux in Eq.
(\ref{prop}) we define the dimensionless flux normalization factor
$N_{s}^{\sigma}$ for the mixed state. Here and in the following $\sigma$ is
the chirality and $s$ the root of an eigenstate. Note that this normalization
is rather arbitrary. We could have also used angular momentum flux
normalization, or fix the amplitude of one component to unity. We believe,
however, that for more general situations with reduced symmetry and many wave
vectors, the present choice is most convenient.

\section{Interface boundary conditions}

\begin{figure}[ptb]%
\centering
\includegraphics[
natheight=2.074700in,
natwidth=4.150200in,
height=1.3493in,
width=2.6822in
]%
{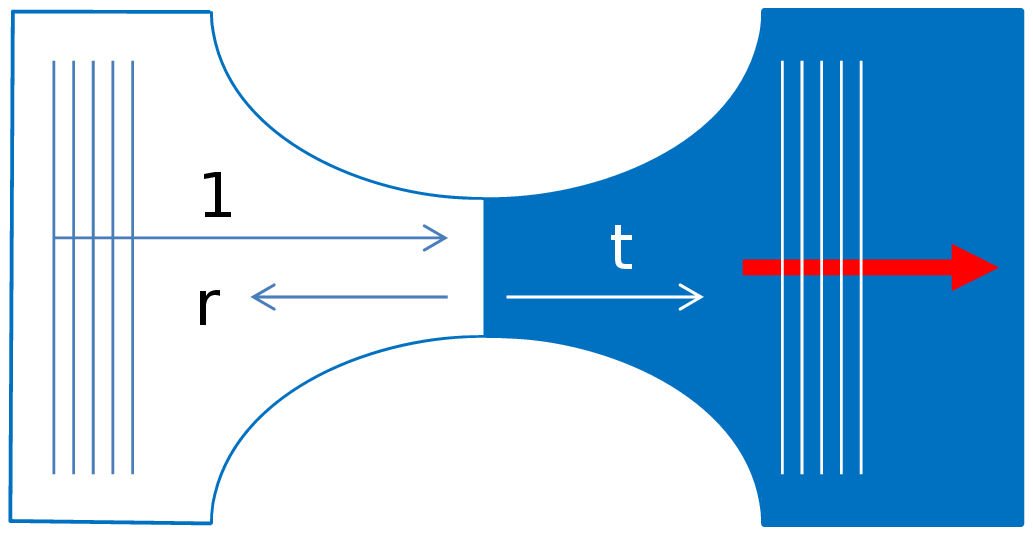}%
\caption{Schematic scattering problem of a phonon reflected and transmitted
at the interface of a ferromagnet. The red arrow is the magnetization ($z$-)
direction.}%
\label{LB}%
\end{figure}

We consider a weakly damped ferromagnetic structure actuated by a
piezoelectric layer (Fig. \ref{LB}) that is excited at a given resonance
frequency $\omega.$ We assume that any reflection vanishes at the end of the
ferromagnet, e.g. by attaching an acoustic absorber \cite{Uchida}. Both
actuator and ferromagnet are thus considered to be reservoirs adiabatically
connected to the scattering region. In and outgoing waves are then all
propagating. We then may disregard negative wave numbers in the ferromagnet as
well as the scattering coefficients $\mathfrak{r}^{\prime}$ and $\mathfrak{t}%
^{\prime}$, the reflection and transmission coefficients of waves from the
magnetic side. On the left side we have incoming circularly polarized phonons
with chirality $\sigma$ that is conserved when reflected at a flat interface
to a ferromagnet with magnetization along the propagation direction%
$\mathcal{\ }$%
\begin{equation}
\chi_{L}^{\sigma}\left(  z,t\right)  =\frac{N_{p}^{L}}{k_{L}}\left[  \left(
\begin{array}
[c]{c}%
0\\
1
\end{array}
\right)  e^{ik_{L}z}+\left(
\begin{array}
[c]{c}%
0\\
\mathfrak{r}_{L}^{\sigma}%
\end{array}
\right)  e^{-ik_{L}z}\right]  e^{i\omega t},\;
\end{equation}
where $N_{p}^{L}=\left(  4\alpha_{L}^{3}\rho_{L}\right)  ^{-1/4},$
$k_{L}=v_{L}\omega=\sqrt{\alpha_{L}/\rho_{L}}\omega$ in terms of the acoustic
parameters of the actuator and $\mathfrak{r}_{L}^{\sigma}$ is the reflection
coefficient determined below. This state carries an energy flux of $F_{\sigma
}=F_{0}^{\sigma}\left(  1-\left\vert \mathfrak{r}_{L}^{\sigma}\right\vert
^{2}\right)  ,$ where $F_{0}^{\sigma}$ is the actuator power density. On the
right side we can scatter into the two mixed eigenstates at the same frequency
with transmission coefficients $\mathfrak{t}.$ The axial symmetry prevents
mixing between states with different polarizations and
\begin{equation}
\chi_{R}^{\sigma}\left(  z,t\right)  =e^{i\omega t}\sum_{s}\mathfrak{t}%
_{s}^{\sigma}\left(  \omega\right)  \left(
\begin{array}
[c]{c}%
m_{s}^{\sigma}\\
r_{s}^{\sigma}%
\end{array}
\right)  e^{ik_{s}^{\sigma}z} \label{amp}%
\end{equation}
where the magnetic and lattice components are flux-normalized as described
above. Energy conservation dictates that
\begin{equation}
\frac{F}{F_{0}}=1-\left\vert \mathfrak{r}_{L}^{\sigma}\right\vert ^{2}%
=\sum_{s}\left\vert \mathfrak{t}_{s}^{\sigma}\right\vert ^{2} \label{flx}%
\end{equation}
which reflects the unitarity of the scattering matrix composed by
$\mathfrak{r}$ and $\mathfrak{t}$ (as well as by $\mathfrak{r}^{\prime}$ and
$\mathfrak{t}^{\prime}$).

\begin{figure}[tbh]
\centering
\includegraphics[scale=0.3]{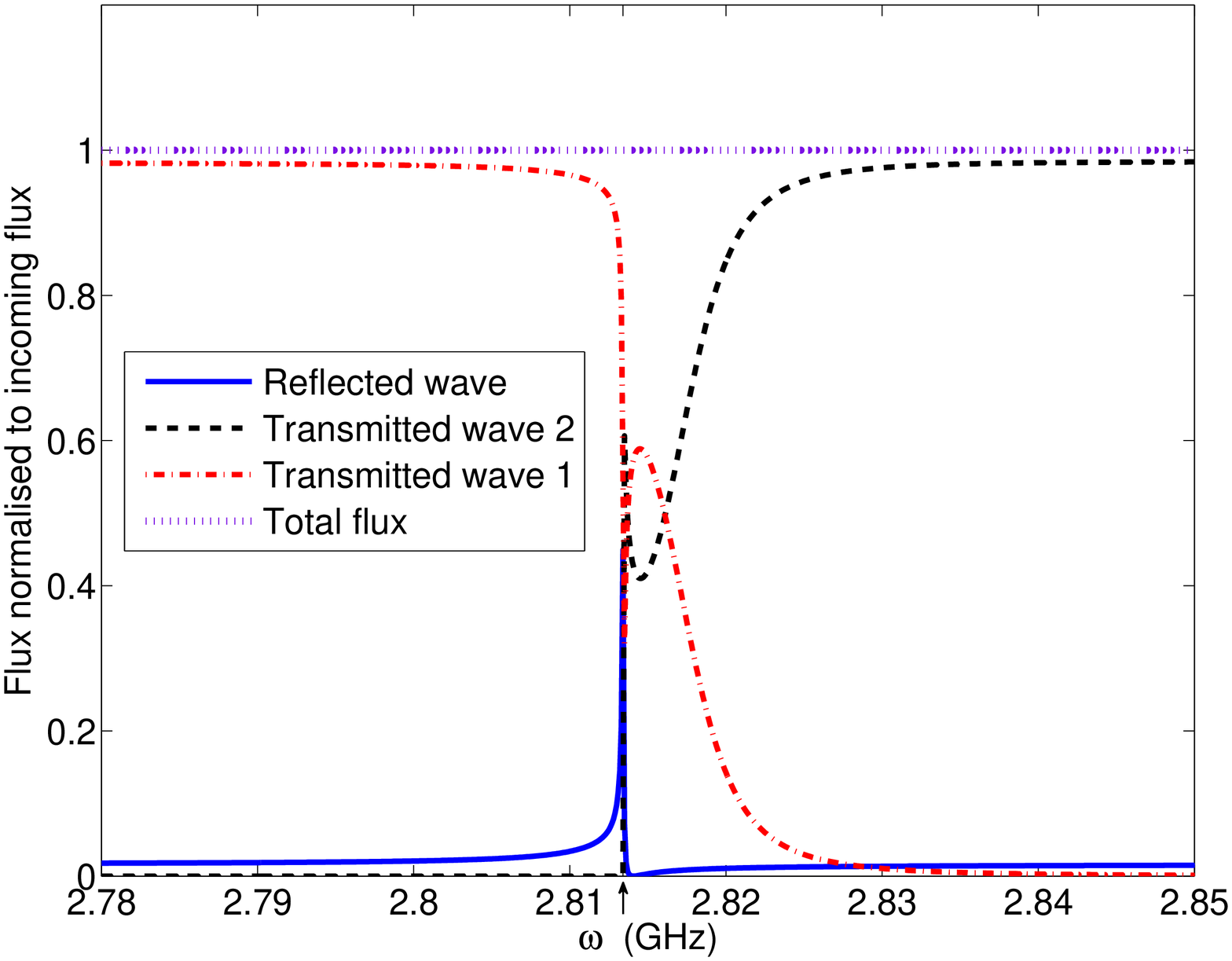} \caption{Normalized flux components
$\left\vert \mathfrak{r}_{L}^{\sigma}\right\vert ^{2},\left\vert
\mathfrak{t}_{1}^{\sigma}\right\vert ^{2},$ and $\left\vert \mathfrak{t}%
_{2}^{\sigma}\right\vert ^{2}$ as a function of frequency close to the
anticrossing between magnon and phonon modes. The labels of the transmitted
waves are consistent with the labeling of the wave numbers in Fig.
\ref{dispall} (a) and (c). Flux conservation $\left\vert \mathfrak{r}%
_{L}^{\sigma}\right\vert ^{2}+\left\vert \mathfrak{t}_{1}^{\sigma}\right\vert
^{2}+\left\vert \mathfrak{t}_{2}^{\sigma}\right\vert ^{2}=1$ is demonstrated.
The FMR resonance frequency $\omega_{0}$ is indicated by the arrow on the
abscissa.}%
\label{transrefl}%
\end{figure}

At the interface $z=0$ we demand continuity of the lattice $R^{\sigma}\left(
0^{-}\right)  =R^{\sigma}\left(  0^{+}\right)  ,$ which leads to%
\begin{equation}
\frac{N_{p}^{L}}{k_{L}}\left(  1+\mathfrak{r}_{L}^{\sigma}\right)  =\sum
_{s}\mathfrak{t}_{s}^{\sigma}r_{s}^{\sigma}.
\end{equation}
Continuity of the stress or energy current at the interface $z=0$, $\bar
{F}\left(  0^{-}\right)  =\bar{F}\left(  0^{+}\right)  ,$ leads to the
boundary condition
\begin{equation}
\frac{b_{2}}{M_{0}}\sum_{s}\mathfrak{t}_{s}^{\sigma}m_{s}^{\sigma}=i\alpha
_{L}\frac{N_{p}^{L}}{k_{L}}\left(  1-\mathfrak{r}_{k_{L}}\right)  -i\alpha
_{R}\sum_{s}\mathfrak{t}_{s}^{\sigma}r_{s}^{\sigma}k_{s}^{\sigma}%
\end{equation}
Integrating the equation of motion over the interface leads to free boundary
condition for the magnetization $\partial m^{\sigma}/\partial x\left(
0^{+}\right)  =0,$ which implies that the energy and angular momentum carried
by spin waves vanish at the boundary, leading to a relation for the
transmission coefficients%
\begin{equation}
\sum_{s}\mathfrak{t}_{s}^{\sigma}m_{s}^{\sigma}k_{s}^{\sigma}=0
\end{equation}
We have now three linear equations with three unknown variables for a choice
of chirality, \textit{viz}. $\mathfrak{t}_{1}^{\sigma},\,\mathfrak{t}%
_{2}^{\sigma}$ and $\mathfrak{r}_{L}^{\sigma}.\ $We can easily derive the
coefficients as analytic functions of $\omega$ and constituting parameters,
but the expressions are lengthy, hence, not given here. We used parameters of
gadolinium gallium garnet (GGG) for the actuator: $\rho_{L}%
=7085\,\mathrm{kg/m^{3}},\,\alpha_{L}=9.0\times10^{10}\,\mathrm{Pa}$, which
implies small but finite acoustic mismatch. In Fig. \ref{transrefl}, the flux
components at the interface are plotted as a function of frequency for the
parameters used in the dispersion relations (Fig. \ref{dispall}).
Far from the (anti) crossing we observe weak reflection of the
incoming sound wave, which is a consequence of the good acoustic impedance
matching assumed here. Transmission into the phonon-like root $s$ that changes
signs between low and high frequencies dominates. The transmission and
reflection probabilities look complicated because the spin wave dispersion is
so flat; the anticrossing overlaps with the FMR frequency $\omega_{0}$ for
spin wave excitation, indicated by the arrow on the abscissa. This is so
because the exchange energy at the anticrossing is insignificant as compared
to the Zeeman energy. In the neighborhood of the anticrossing we observe the
typical mode conversion plus an additional contribution to the back reflection
of acoustic energy into the actuator.

\section{Detection of magnetization dynamics by spin pumping}

\begin{figure}[tbh]
\centering
\includegraphics[scale=0.3]{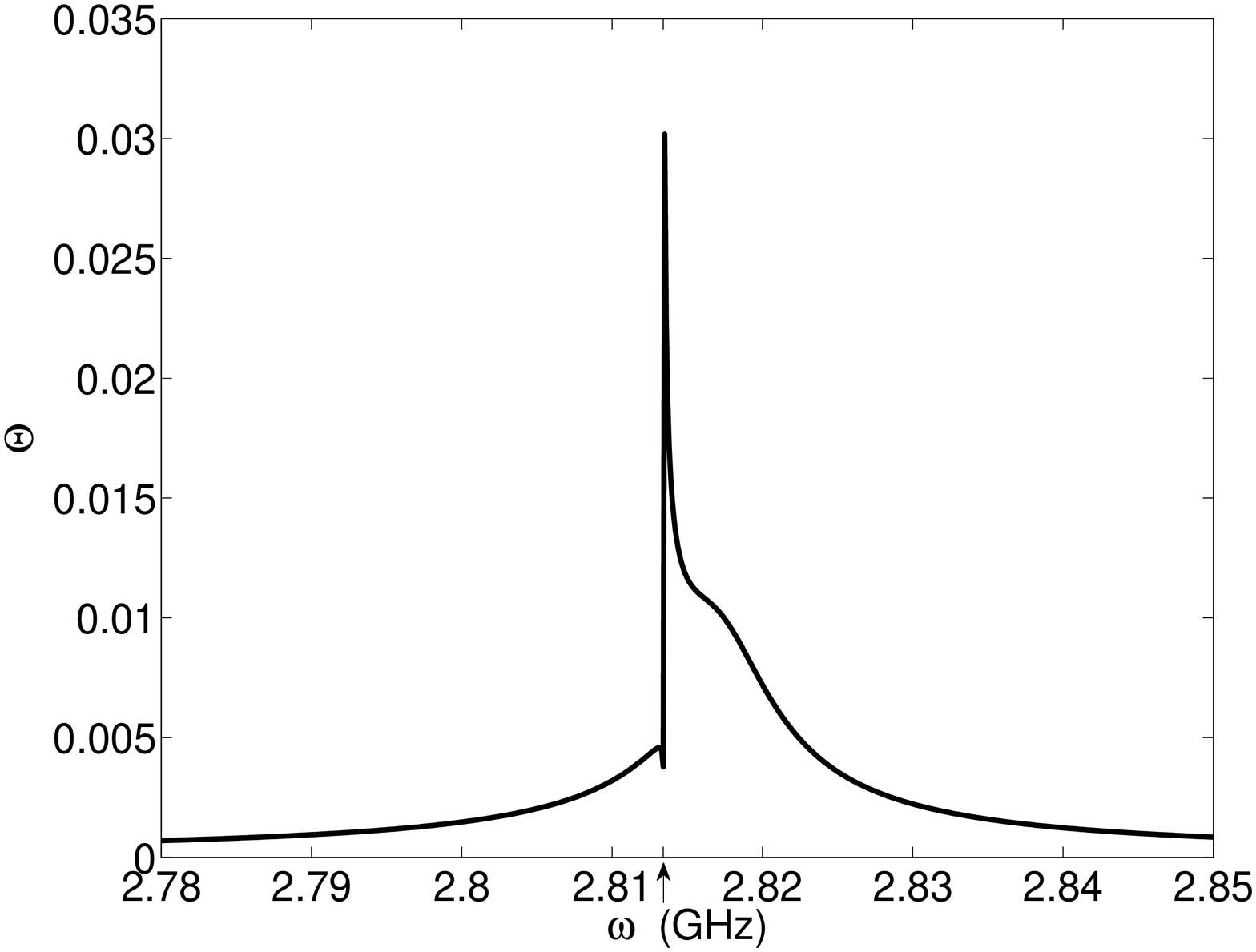} \caption{$\Theta$ (Eq. (\ref{theta2}))
\textit{vs}. frequency of the injected sound wave for unit injected acoustic
energy $F_{0}=1\,\mathrm{J/}\left(  \mathrm{m}^{2}\mathrm{s}\right)  $
($\Theta\sim\sqrt{F_{0}}$) . $\Theta$ is the average precession cone angle and
a metric for the efficiency of acoustic excitation of the magnetic degree of
freedom. The FMR resonance frequency $\omega_{0}$ is indicated by the arrow on
the abscissa.}%
\label{eeta}%
\end{figure}

Uchida \textit{et al}. \cite{Uchida} and Weiler \textit{et al.}
\cite{Goennenwein} detected the acoustically induced magnetization dynamics by
the spin current pumped \cite{pump} into a thin layer of Pt with a significant
spin Hall angle $\theta_{\mathrm{SH}}$ \cite{Hoffmann}. In terms of the spin
mixing conductance $g_{r}^{\uparrow\downarrow}$ the magnitude and polarization
of the spin current reads \cite{pump}
\begin{align}
\mathbf{I}_{s}^{\mathrm{pump}}  &  =\frac{\hbar}{4\pi}\frac{g_{r}%
^{\uparrow\downarrow}}{M_{0}^{2}}\mathbf{M}\times\frac{d\mathbf{M}}%
{dt}\label{spinemission}\\
&  =\frac{\hbar}{4\pi}\frac{g_{r}^{\uparrow\downarrow}}{M_{0}^{2}}\left(
\begin{array}
[c]{c}%
M_{0}\dot{M}_{x}\mathbf{\hat{y}}-M_{0}\dot{M}_{y}\mathbf{\hat{x}}\\
+\left(  M_{x}\dot{M}_{y}-M_{y}\dot{M}_{x}\right)  \mathbf{\hat{z}}%
\end{array}
\right)  .
\end{align}
For $M_{x}^{2}+M_{y}^{2}\ll M_{0}^{2}$ the average cone angle
\begin{align}
\Theta &  =\sqrt{\left\langle M_{x}^{2}+M_{y}^{2}\right\rangle _{z,t}}%
/M_{0}\label{theta2}\\
&  =2\sqrt{\sum_{s}\left(  \left\vert t_{s}^{+}\right\vert ^{2}\left\vert
m_{s}^{+}\right\vert ^{2}+\left\vert t_{s}^{-}\right\vert ^{2}\left\vert
m_{s}^{-}\right\vert ^{2}\right)  }/M_{0}%
\end{align}
is a convenient metric for the excitation of the magnetic degree of freedom
per unit acoustic energy flux. The angular brackets indicate a time and
position average over interference fringes that are an artifact of the
one-dimensional and monochromatic approximations. $\Theta\left(
\omega\right)  $ is plotted in Fig. \ref{eeta}. The qualitative features can
be understood in terms of competition between magnetic character and
excitation efficiency of the eigenmodes. As the frequency approaches the
anticrossing from below, the increasing magnetic character of the propagating
normal mode leads to an increasing $\Theta$. Just below the FMR frequency, the
normal mode becomes flat, corresponding to a small group velocity which
reduces mode excitation efficiency (see Fig. \ref{transrefl}). Just above the
FMR frequency, a new propagating normal mode becomes available which restores
a relatively efficient mode excitation. The highest $\Theta$ is achieved
close to the crossing at which the acoustic and magnetic modes are fully
mixed. Finite dissipation, finite quality factor of the actuator, and disorder
disregarded here will broaden the sharp features and reduce the magnetization
amplitude, but the effects are believed to be minor for high quality YIG and
proper actuator design.

When the Hall contacts of the Pt layer are short circuited, the spin current
induces a Hall charge current in the Pt layer with direction and magnitude
\cite{Hoffmann}
\begin{equation}
\mathbf{I}_{c}=\frac{2e}{\hbar}\theta_{\mathrm{SH}}\mathbf{\hat{z}}%
\times\mathbf{I}_{s}^{\mathrm{pump}}%
\end{equation}
with
\begin{align}
\left(  \mathbf{I}_{c}\right)  _{x,y} &  =-\frac{eg_{r}^{\uparrow\downarrow}%
}{2\pi}\theta_{SH}\frac{\dot{M}_{x,y}}{M_{0}}\\
\sqrt{\left\langle \left\vert \left(  \mathbf{I}_{c}\right)  _{x}\right\vert
^{2}+\left\vert \left(  \mathbf{I}_{c}\right)  _{y}\right\vert ^{2}%
\right\rangle } &  =\frac{eg_{r}^{\uparrow\downarrow}\theta_{SH}}{2\pi M_{0}%
}\sqrt{\left\langle (\dot{M}_{x})^{2}\right\rangle +\left\langle (\dot{M}%
_{y})^{2}\right\rangle }\\
&  =\frac{eg_{r}^{\uparrow\downarrow}\theta_{SH}\omega\Theta}{2\pi}.
\end{align}
We see that the DC spin current component does not generate an inverse spin
Hall effect in this configuration, but a pure AC inverse spin Hall effect is
expected \cite{Hujun}. The currents in the $x$ and $y$-direction oscillate at
the frequency of the actuating ultrasound with amplitude $\sim\Theta,$ in
contrast to the DC inverse spin Hall effect that scales with $\Theta^{2}$ and
is orders of magnitude smaller for small acoustic energy fluxes. The FMR
generated AC inverse spin Hall effect is difficult to observe since the Pt
layer is directly subjected to rf radiation, which causes strong
electrodynamic artifacts at the resonance frequency. The acoustically
generated AC inverse spin Hall effect does not suffer from this disadvantage
when the piezoelectric material is spatially separated from the magnetic material.

\section{Conclusions}

We computed the acoustically stimulated magnetization dynamics and the
associated spin-pumping induced ac inverse spin Hall effect for a symmetric
configuration of transverse acoustic waves polarized normal to the
magnetization direction. The theory can be extended to include longitudinal
phonons (pressure waves) and arbitrary magnetization direction, finite quality
factor of the actuator, finite magnetizations damping, multilayered
structures, diffuse scattering by disorder, etc., if necessary. In principle
this formalism can be extended as well to obtain spin and heat transport
through arbitrary structures under a temperature difference between the
reservoirs by including incoming and outgoing waves with all wave vectors.

\section*{Acknowledgement}

The authors thank Peng Yan and Sebastian G\"{o}nnenwein for useful
discussions. This work was supported by the FOM Foundation, Marie Curie ITN
Spinicur, Reimei program of the Japan Atomic Energy Agency, EU-ICT-7
\textquotedblleft MACALO\textquotedblright, the ICC-IMR, DFG Priority
Programme 1538 \textquotedblleft Spin-Caloric Transport\textquotedblright, and
Grand-in-Aid for Scientific Research A (Kakenhi) 25247056.

\end{document}